# Inhomogeneous Superconducting State and intrinsic $T_c$ : Near Room Temperature   Superconductivity in the Cuprates.

Vladimir Z.Kresin* and Stuart A.Wolf**

* Lawrence Berkeley National Laboratory, University of California at Berkeley, CA 94720

** Departments of Physics and Materials Sciences, University of Virginia Charlottesville, VA  22904

Doped cuprates are inhomogeneous superconductors. The concept of an intrinsic critical temperature , $T_c^{intr.} \equiv T_c^*$, whose value greatly exceeds that for th resistive  $T_c^{res.} \equiv T_c$, is supported by a number of experimental studies, including those performed recently. These data are discussed in this review.  The anomalous  diamagnetism observed  at $T_c^{res.} \equiv <T< T_c^*$ is a manifestation of the presence of superconducting clusters embedded into a  normal metallic matrix. The value of intrinsic critical temperature  in some cuprates reaches the value which is close to room temperature. The a.c. properties of such inhomogeneou systems are discussed.

**Introduction**. At present, searches for superconducting systems with higher value of $T_c$ with the goal to achieve the room temperature superconductivity has attracted a lot of attention ; this is due to the progress in the study of new materials. The high $T_c$ cuprates, discovered in [1] with following development [2,3], despite an intensive study, continue to be a puzzling family of compounds. As we know, the maximum value of temperature of their transition to the macroscopic superconducting state is equal to $T_c \approx 135K$ ( under pressure $T_c = 164K$ ). We would like to stress in this paper that in reality the value of their intrinsic critical temperature in the cuprates, $T_c^{intr.} \equiv T_c^*$, is much higher. For example, for the underdoped YBCO as well as for underdoped Bi-2212 compounds ($T_c \approx 85K$) the value of $T_c^* \approx 200\text{-}250K$. For the underdoped LaSrCuO ($T_c \approx 24K$) the value of its intrinsic critical temperature $T_c^* \approx 80K$. These values are, indeed, much higher than the resistive $T_c$. These large values are due to the peculiar nature of the transition into the superconducting state in doped systems, especially in the underdoped region. Below we will
discuss this question in more detail and propose some experiments which will support our point of view, and, in addition, will lead to interesting applications.

As we know, the part of the phase diagram above $T_c$ forms the so-called "pseudogap" region. This pseudogap state is particularly obvious for the underdoped region of the phase diagram. Our analysis of this state was described in [4-7]. Here we describe recent important results, obtained by several experimental groups and propose some additional experiments. We focus on the concept of intrinsic critical temperature, $T_c^*$, and emphasize that the cuprates are almost room temperature superconductors.

The key issue which is still controversial is related to the nature of the "pseudogap" state. The main question is whether this state is a normal metal or superconductor. At first sight, the observation of the finite resistance provides the answer to this question. Nevertheless, in reality the situation is more complicated. The first complication arises from the observation of the energy gap spectrum ( this is reflected in the name of the phenomenon, although the name is rather misleading).

In connection with this let us note that the presence of the energy gap is not a sufficient factor to resolve the question. Indeed, the energy gap could correspond to the superconducting state, but also, e.g., to the CDW or SDW states of the normal metal. It is worth noting, however, that the smooth evolution of the gap above $T_c$ into that below $T_c$ (see,e.g. [ 8 ]) is a serious indication that the state at $T > T_c$ is related to superconductivity.

**Diamagnetism above Tc** . However, there is a more fundamental observation. Namely, several experimental groups using different techniques observe a strong and temperature dependent diamagnetic effect above $T_c$. The first observations were discussed by us in [ 4 ] , see also [ 7 ]. According to the concept we introduced , the coexistence of normal resistance and an anomalous diamagnetism (Meissner effect) could be explained by an inhomogeneity in the system, namely, by the presence of superconducting " islands" in the normal metallic matrix.

During the last several years there have been new important observations of the diamagnetic phenomenon which we will discuss below. They were not described in our review [ 7 ].

But at first let us mention some previous experimental observations [ 9,10 ] , see also [ 7 ]. Scanning SQUID microscopy was used to study the underdoped $La_{2-x}Sr_xCuO_4$ compound [ 9 ]. This method allows one to create a "magnetic" map. A peculiar inhomogeneous picture with diamagnetic regions has been observed ( see [ 9 ] and Fig.1 in [7 ]) . The critical temperature of the underdoped LSCO films was: $T_C \approx$ 18K. As for the observed diamagnetism, the film contains diamagnetic domains and their presence persists up to 80K (!). The total size of the diamagnetic regions grows as the temperature is decreased .The diamagnetic response appears to be strongly temperature dependent; this is a very unusual feature of these materials.

Previously a strong diamagnetic response has been also observed [10 ]) by using the torque magnetometry technique for the overdoped $Tl_2Ba_2CaO_{+\delta}$ compound above $T_C \approx$ 15K. Like LSCO, the diamagnetic moment was also strongly temperature dependent .

Let us now discuss new data. A very interesting experiment was described in [ 11 ]. The authors [11 ] use a technique entirely different from that employed in [9,10 ] , namely they measured the µsR relaxation. This method allows one to measure directly the local magnetic field. As we know ,the µsR spectroscopy (µsR stands for muon spin relaxation) is the experimental technique based on the implantation of spin polarized muons into the sample (see, e.g. reviews [12,13] In many respects this technique is similar to the electronic paramagnetic resonance (EPR) and , even more, to the nuclear magnetic resonance (NMR) methods. The method allows one to obtain the information about the local magnetic field, since it affects the spin precession. The study [11 ] is of special interest, because the µsR method , unlike the STM studies is a *bulk* measurement. According to the data [11], one observes a strong magnetic response above $T_c$ . More specifically, the relaxation process is described by the function $G(t)=\exp[-(\Lambda t)^\beta]$, where $\beta$=const. The authors [ 11 ] studied the YBCO and LSCO samples. They compare the relaxation with the Ag samples which is not a superconducting metal. If the state at T>$T_c$ is an usual normal metal, one should expect the behavior similar to that for Ag. However, according to [ 11 ] ,the behavior of $\Lambda$ for LSCO and YBCO at T>$T_c$ is different ; the quantity $\Lambda$-$\Lambda_{ag}$ strongly depends on temperature , and this corresponds to the presence of magnetic regions above $T_c$.

**Inhomogeneity: coexistence of diamagnetism and finite resistance. Superconducting "islands".** Orbital diamagnetism ( Meissner effect) is the most fundamental property of the superconducting state. As we know, a normal metal does not display any noticeable diamagnetic effect (except a very small Landau term). It is important to stress that the diamagnetism observed in [ 9.11] can not be explained by fluctuations, because they are important only near $T_c$ , whereas the observed diamagnetism extends to much higher temperatures. This point about fluctuations is supported by a more detailed analysis carried out in [ 14.15 ].

Therefore, we are dealing with a serious challenge. Namely, one should explain the coexistence of finite resistance and anomalous diamagnetism. As was mentioned above, according to our theory, the "pseudogap" state is cause by intrinsic inhomogeneity of the metallic phase. Namely, above $T_c$ the system (in the region $T_c < T < T_c^*$) is characterized by the presence of superconducting regions embedded into a normal metallic matrix. In other words, the inhomogeneit leads to a local value of critical temperature that is spacially dependent:

$T_c^{loc.} \equiv T_c^{loc.}(r)$.

Note that as a whole the metallic phase (including the superconducting "islands") coexists with the insulating regions (see below for a discussion about the energy scales).

The coexistence of a normal metallic matrix and superconducting regions ("islands") leads to the proximity effect playing a very important role. This effect was explicitly taken into account in our calculations of the diamagnetic response, the energy spectrum and a.c. properties [4-7]. Note also that because of the proximity effect the size of the superconducting regions should be of order or larger than the coherence length; otherwise, the superconducting state will be destroyed by the proximity contact with the normal matrix.

The superconducting regions ("islands") that appear at $T < T_c^*$; remember $T_c^*$ is a characteristic temperature that can be labeled as an intrinsic $T_c$. Each such region has its own phase. The consequent decrease in T below $T_c^*$ leads to an increase in the size of such regions and also an increase in the number of " islands". Finally, at $T = T_c$, we are dealing with the percolative transition into a macroscopic dissipationless superconducting state with phase coherence throughout the sample.

**Inhomogeneous and superconducting state above the resistive Tc: new STM and ARPES data.** Recently several new experimental studies have been carried out (e.g., [ 11,16-18,20-23 ], and they are in a total agreement with our approach. The μsR relaxation experiment [ 11 ] was discussed above,

The important study was performed using scanning tunneling microscopy (STM) [ 16,17 ]. The measurements were performed on the $Bi_2Sr_2CaCu_2O_{8+\delta}$ samples at finite temperature, so that the energy spectrum was measured above $T_c$, in the "pseudogap" region. The measurements described in [16,17], have provided crucial information about local values of the gap and its evolution with temperature. The study of this evolution has led to the conclusion that the observed gap spectrum, indeed, corresponds to superconducting pairing. It is essential to realize that the distribution of gaps turns out to be strongly inhomogeneous. As a whole, the presence of pairing persists for temperatures which greatly exceed those of $T_c$, especially for the underdoped samples. The characteristic temperature pairing to appear is denoted in [16,17 ] as $T_p$; this quantity was denoted above as $T_c^*$, so that $T_p \equiv T_c^*$. For example, for the sample with the doping level x≈0.1 ($T_c$≈75K) the pairing persists up to $T_c^*$≈180K. It is interesting also that from the measurements [ 16,17 ] one can determine also the local values of the gap $\Delta_{loc.}$ and $T_p^{loc.}$. One can conclude that the ratio2 $\Delta_{loc}$ / $T_p^{loc}$ appears to be equal to $2\Delta_{loc} / T_p^{loc} \approx 7.5$. Such a large value of this ratio greatly exceed that for the usual superconductors (according to the BCS theory $2\Delta /T_c \approx 3.52$) and corresponds to very strong electron-phonon coupling ( see, e.g., [ 19 ]).

As a whole, the study [ 16,17 ] provides a strong experimental evidence for the existing of the inhomogeneous superconducting state above $T_c$, in the "pseudogap" region $T_c<T<T_c^*$.

As a whole, the studies [16,17] provide strong experimental evidence for the existence of the inhomogeneous superconducting state above $T_c$, in the "pseudogap" region $T_c<T<T_c^*$. The presence of pairing above $T_c$ has been demonstrated also using angle resolved photoemission (ARPES). The analysis of energy-momentum dispersion shows that near the Fermi arcs which correspond to the normal metal there is a region which corresponds to dispersion typical for a gapped superconducting state. This study also provide support to the picture of the coexistence of a normal metallic matrix and superconducting clusters embedded in this matrix.

We mentioned above the measurements of the ratio $2\Delta / T_c^*$ [16,17]. In connection with this one should recall that, according to the BCS theory, the value of energy gap is proportional to the critical temperature; for the superconductors with strong coupling the relation is more complex ( see, e.g.,[19]),but still there is a direct one-to-one correspondence between these quantities. For the cuprates one should distinguish the intrinsic $T_c^*$ and $T_c \equiv T_c^{res.}$.

It is clear that, according to our approach, the pairing gap is related to $T_c^*$, since the Cooper pairing occurs at first at this temperature. That's why the changes in the value of $T_c^*$ lead to corresponding changes in the gap value. As for $T_c \equiv T_c^{res,}$, this temperature describes the percolation transition to the macroscopic superconducting state and is not related directly to the pairing interaction. Correspondingly, the changes in Tc should not affect the gap value. It is interesting that precisely this picture has been observed in [20]. The values of $T_c^*$ and $T_c$ were changed independently (e.g., the value of $T_c$ was modified by Zn substitution). Indeed, the value of the gap measured using ARPES was sensitive to the changes in $T_c^*$, but was not affected by changes in the resistive $T_c$.

The successful development of the ARPES and STM methods has resulted in various measurements of the energy gap spectrum as we have described. Recently, several groups claimed detailed measurements related to the question as to whether we are dealing with one or two energy gaps above $T_c$. According to some studies (e.g. [21]) there is a single pairing gap above $T_c$, which continuously changes into the gap below $T_c$. a similar transition also occurs in the overdoped region [22]. Other studies (e.g. [23]) show the presence of two gaps above $T_c$, one in the nodal direction (pairing gap) and another one in the antonodal direction that is not related to the superconducting pairing (it could correspond to CDW ,SDW, etc).

From our point of view it is essential that all data indicate the presence of a superconducting gap above $T_c$. Some authors use a rather vague terminology "precursor pairing" implying some type of pairing without the phase coherence which arises at $T_c$. we think that this "precursor pairing" which implies no local phase coherence contradicts the observation of the "giant" Josephson effect above $T_c$ [24,6], see below. Our scenario is based on a picture of superconducting "islands" embedded into normal metallic matrix, and each "island' has its own phase. The percolation transition at $T_c$ leads to the formation of a single macroscopic phase having a single value..

**Energy scales.** According to [4-7], one should distinguish three energy scales : $T_c, T_c^*, T^*$, so that $T_c < T_c^* < T^*$. At $T<T^*$ one can observe the coexistence of the metallic and insulating phases (phase separation). The concept of phase separation was introduced in [25] and then developed in many papers (see, e.g., [26]). In principle, the region $T<T^*$ can display a spectrum reflecting an energy gap, but until $T> T_c^*$ it is not related to the pairing. For example, the CDW (or SDW) gap could be observed in this region. One can mention here that the presence of the chain structure in the YBCO compound leads to the appearance of a corresponding CDW gap spectrum.

As was emphasized above, $T_c^*$ is a very important parameter: below $T_c^*$ the superconducting regions ("islands") appear and this leads to the formation of pairing gaps. The transition at $T_c^*$ is manifested by the appearance of a diamagnetic signal. The superconducting "islands" are embedded in a normal metallic matrix. Such a coexistence of superconducting clusters and the normal metallic matrix is caused by the inhomogeneity of the system. Inhomogeneity is reflected , e.g., in the spatial distribution of the energy gaps [16 ]. Each "island" has its own phase, so that the phase coherence over a whole sample does not exists in this temperature region. The inhomogeneity means that the value of the critical temperature is spatially dependent. The consequent decrease in T leads to an increase in the area occupied by superconducting clusters ; thus we are dealing with the percolation scenario. Eventually at $T=T_c$ the macroscopic superconducting phase is formed ( percolation transition), and below $T_c$ one can observe the dissipationless transport (R=0) and macroscopic phase coherence.

    Let us define the characteristic temperature $T_c^*$ which corresponds to the Meissner transition as the " intrinsic critical temperature". Its value could be close to room temperature. For example, for YBCO the value of $T_c^*$ is 250K [18].

    It is important to realize that the value of $T_c^*$ depends on the doping level. Its highest value corresponds to the underdoped region. A sample with optimum doping has an almost homogeneous structure. As a result, an increase in doping towards the optimum value leads to the highest value of the resistive critical temperature, $T_c$, however this increase in doping leads to decrease in the value of $T_c^*$, so that near the optimum doping $T_c^*$ is close to $T_c$. Such a decrease of the intrinsic critical temperature $T_c^*$ is an key feature of the doping; it allows us to describe the unusual isotope effect observed in the "pseudogap" state (see below).

**Origin of inhomogeneity**. The question is: what is the origin of inhomogeneity? In principle, there are two possible scenarios:

- inhomogeneous charge distribution, and
- non-uniform distribution of pair-breakers

In our papers [ 4-7 ] we consider both these pictures. Let us note that, according to the NMR data [27], the carrier density in the cuprates is uniform. Then we are dealing with a picture of a non-uniform distribution of pair-breakers. The pair-breaking can be provided by doping and the statistical nature of doping correlates with the inhomogeneous distribution of dopants.

As we know, the pair-breaking effect [ 28 ] can be caused by localized magnetic moments. For the D-wave scenario even non-magnetic defects appear to be pair-breakers. For the cuprates which are doped materials the dopants play the role of such defects. It is important that pair-breaking leads not only to depression of the energy gap parameter, but also to depression in the value of the critical temperature [28], see also [7]. The statistical nature of doping leads to an inhomogeneous superconducting state, so that $T_c^{loc.} \equiv T_c^{loc.}(r)$.

**Isotope effect.** According to an interesting experiment [29], the "pseudogap' state is characterized by a strong isotope effect, that is, by a large shift in $T_c^*$ caused by isotope substitution. The authors [29] studied the $HoBa_2Cu_4O_8$ compound with a value of $T_c^* \approx 170K$. The isotopic substitution $^{16}O \rightarrow ^{18}O$ leads to a drastic change in the value of $T_c^*$. One can observe the following change : $T_c^* \approx 170K \rightarrow T_c^* \approx 220K!$. We are dealing with a giant isotope effect ; in addition, its value is negative.

Such an isotopic dependence of $T_c^*$ is also a consequence of the presence of the superconducting regions. It is worth mentioning that the doping level greatly affects the value of the isotope coefficient for the usual resistive $T_c$. This dependence was explained in our paper [30] , see also [ 7, 19].

The cuprates are doped materials and the charge transfer (doping) plays a very important role. To describe this process, one should take into account the dynamic polaronic effect. More specifically, the dynamics of oxygen ions is described by a double-well potential. In other words, such an ion has two close equilibrium positions separated by a barrier. Such a double-well structure was observed experimentally using of X-rays absorption in [31]. As a result, the charge transfer between the reservoir and the CuO layer which occurs through such an oxygen ion (e.g. an apical oxygen in YBCO) can be visualized as a two-step process. Initially the carrier makes the transition from the chain site to the apical oxygen O(4), then the apical oxygen transfers to another minimum (tunneling), and this is finally followed by the transition of the carriers to the CuO plane. The tunneling step is affected by the isotopic substitution.

The total wave function can be written in the form:

$$\Psi(\vec{r},\vec{R}) = a\psi_1(\vec{r},\vec{R})\Phi_1(\vec{R}) + b\psi_2(\vec{r},\vec{R})\Phi_2(\vec{R})$$

Here

$$\psi_i(\vec{r},\vec{R}), \Phi_i(\vec{R})$$

(1)

are the electronic and vibrational wave functions, corresponding to the minima, i={1.2}. $\{\vec{r},\vec{R}\}$ are the electronic and ionic coordinates. One can see directly from Eq.(10) that the total wave function can not be written just as the product of the electronic and vibrational wave functions. Contrary to the usual adiabatic scenario, we are dealing with a giant non-adiabatic phenomenon when the electronic and vibrational degrees of freedom can not be separated. An analysis of such a dynamic polaronic effect carried out in [30] (see also [7,19] that leads to the following expression for the isotope coefficient:

$$\alpha = \gamma \frac{n}{T_c} \frac{\partial T_c}{\partial n}$$

(2)

γ=const, n is the carrier concentration.

A recent study [31] contains an interesting analysis of the effect of isotope substitution performed for several families of the cuprates. More specifically, the authors [31] have studied the behavior of the sotope coefficient as a function of the carrier concentration. A whole region of the phase diagram from the optimum doping down to near vanishing superconductivity was analyzed. The analysis shows an excellent agreement between the data and Eq.(2) . The paper [32] has demonstrated that the polaronic effects and their local nature are the key ingredients of physics of high $T_c$ cuprates.

In a similar fashion, for the intrinsic critical temperature $T_c^*$ one can write:

$$\alpha = \gamma \frac{n}{T_c^*} \frac{\partial T_c^*}{\partial n} \qquad (3)$$

Contrary to the usual scenario, the isotope effect described by Eq.(3) is not related to the change in vibrational frequency; it reflects the impact of the isotope substitution upon the charge transfer. It is important to realize that its value can exceed the conventional limit $\alpha = 0.5$ This is in agreement with the observations [29].

We stressed also (see above) that the value of $T_c^*$ decreases with increasing n towards to the optimum doping. As a result, the isotope coefficient has a negative sign (see eq.(3)). This was observed in [29] and could be explained with use of Eq.(3).

**High $T_c$ superconducting state. Proposed experiments.** What are manifestations of the high temperature superconducting state? Of course, the presence of the normal matrix at T>Tc excludes the possibility of observing a state with zero dc resistance (R=0). Speaking of transport properties, one can mention an interesting phenomenon, the so-called "giant" Josephson effect which was observed above $T_c$ in [24] and explained in our paper [5] see also [7]. According to [24], the electrodes and the barrier were formed by $La_{0.85}Sr_{0.15}CuO_4$ films ($T_c \approx 45$ K) and the underdoped compound ($T_c' \approx 25K$) respectively. The measurements were performed at $T_c'<T<35$ K, so that we are dealing with an S-N-S junction. The observation of the Josephson current through the "giant" N layer ($L_N$ up to 200A(!) ) was explained in our paper by the formation of superconducting "islands" inside of the barrier whose temperature was in the region $T_c^*>T>T_c$. However, this experiment requires the presence of macroscopic superconducting electrodes and, therefore, the temperature of the whole system can not exceed the values which corresponds to their macroscopic $T_c$.

Nevertheless, one can propose two experiments which demonstrate the presence of a superconducting state at $T>T_c$ which persists up to $T_c^*$. One of them is related to microwave properties of such a composite consisting of superconducting islands inside a normal matrix. The surface impedance of such a composite should be lower than that of just the normal matrix and should be quite temperature dependent reflecting the temperature dependence of the proportion of superconducting material compared to the total. The frequency dependence of such losses should also reflect the energy gap distribution in this composite. In fact, the largest gap of the highest critical temperature island can be determined using a detailed study of the frequency dependence of the microwave absorbtion. At frequencies above this gap frequency the absorbtion close to but below $T_c^*$ should be the same as the absorbtion above $T_c^*$ where everything is normal. The sensitivity of this composite between $T_c$ and $T_c^*$ can also be used as a microwave sensor.

The second experiment is connected with the possibility to observe the ac Josephson effect and corresponding radiation. Contrary to the dc effect, the observation of the ac radiation does not require the presence of macroscopic superconducting electrodes. Consider the layered underdoped cuprate at $T_c^*>T>T_c$. As was noted above, below $T_c^*$ the sample contains superconducting "islands" embedded in normal metallic matrix. These island can form S-N-S Josephson junctions. It is essential that a decrease in temperature leads to an increase in a number of the junctions, so that at $T \approx (T_c^*-T_c)/2$ the superconducting regions form a percolation path and we are dealing with a tunneling network. This tunneling Josephson network can be biased by the current $j>j_c$ or by applied voltage. This leads to an ac Josephson effect, that is, to the phase becoming time dependent and will radiate. One can expect that the frequency will decrease with a decrease in T. Indeed, if an external voltage V is applied between two planes, the percolation path would consist of S-N-S in–plane junctions and the inter-plane junctions transferring the current between the neighboring planes by the intrinsic Josephson effect [33]. Because of phase locking [34.35], the frequency of the ac radiation is ω=eV/hN, where N is the number junctions. Another possibility is connected with observation of Shapiro steps under the influence of incoming radiation.

**Conclusion.** In usual superconductors the resistive and Meissner transitions occur at the same temperature, $T_c$. The inhomogeneous nature of the cuprates leads to a different scenario. Namely, the Meissner effect which is the most fundamental manifestation of superconducting pairing occurs at $T_c^*$ (intrinsic critical temperature), so that $T_c^* > T_c \equiv T_c^{res.}$ The value of $T_c^*$ in some cuprates is close to room temperature and corresponds to the appearance of superconducting clusters inside of the normal metallic matrix. One can propose ta.c. experiments allowing the determination of the presence of superconductivity (the clusters) at such high temperatures with a potential for their applications.